# SDN On BLE: Controlling Resource Constrained Mesh Networks

Yuri Murillo[1,*], Alessandro Chiumento[1,2,*], Brecht Reynders[1] and Sofie Pollin[1]
[1] Telemic, KU Leuven, Department of Electrical Engineering (ESAT), Leuven, Belgium
[2] Connect Centre, Trinity College Dublin, Ireland
*{yuri.murillo, alessandro.chiumento, brecht.reynders, sofie.pollin}*@esat.kuleuven.be
* denotes equal contribution

*Abstract*—Wireless connectivity for the Internet of Things (IoT) requires combining low latency and high reliability with power and cost constraints. On top of that, diverse applications and end-to-end services with very different performance requirements should ideally share the same network infrastructure. Software defined networks (SDN) were defined to allow this coexistence, but they have seen very limited adoption in IoT systems due to the inherent resource constrained nature of the wireless devices. Additionally, the low-power nature of IoT communication standards does not usually allow for physically separated control and data channels, which is a basic requirement for real-time network slicing. In this work, we propose a SDN implementation for Bluetooth Low Energy (BLE), which has become the de-facto technology for IoT applications. The proposed BLE mesh structure is able to use different protocols for different services, ensuring a clear separation between control and data channels while still being sent over the wireless medium. A proof-of-concept for the proposed SDN implementation is given in a real BLE mesh testbed, where measurements show how the system is able to automatically detect and recover from network congestion by identifying the nodes responsible of such situation and reconfiguring their parameters over the air.

## I. INTRODUCTION

Internet of things (IoT) networks consist of low-power sensor nodes with limited computation capabilities, reduced energy and memory, distributed more or less densely over areas to be monitored. Such networks can be deployed at specific pre-planned locations or simply consist of randomly placed and even mobile nodes, thus complicating direct control of them and increasing data routing complexity [1].

The resource constrained nature of the wireless nodes composing the IoT, and Wireless Sensors Networks (WSNs) in general, represents a foundational limitation to the integration of WSNs into mission critical IoT systems in which networks have to gather, distribute and process large amounts of data with reliability and latency constraints [2]. Providing high reliability entails counteracting efficiently network and application dynamics minimising information loss. At the same time, IoT networks require to function for a long time with minimal energy expenditure and control message overhead, thus **adaptive management protocols which scale with dynamic network performance** (due to varying channel quality and/or topology) **are necessary**.

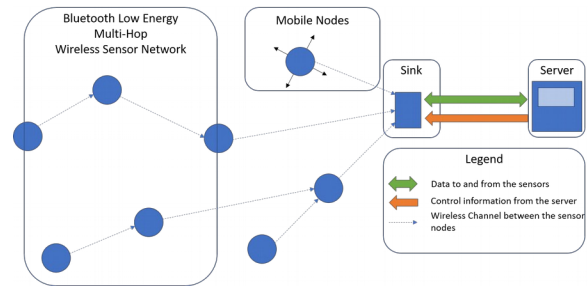

Fig. 1: Bringing SDN to BLE networks is complicated by the resource constrained nature of the nodes and the presence of only the wireless medium for data and control messages.

The Software Defined Networking (SDN) paradigm aims to provide better, more accurate and flexible solutions to the networking and node resource management problems by 1) separating data and control planes and 2) providing efficient centralised control by exploiting the end-to-end properties of the network rather than only local behaviour [3].

Applying the SDN paradigm to WSN is advantageous for many reasons. First, separation of control and data logic leaves wireless sensors as mere data-forwarding devices and pushes the network management to a logically centralised controller, enabling the presence of concurrent applications on a single network. This eases greatly the energy constraints of the network as the most power intensive tasks are then removed from the physical node [4]. Such tasks may include decisions on network topology, routing, data aggregation and storage [3]. Second, a logically centralised controller enables a global view of network performance. By using dedicated APIs (Application Program Interfaces) to communicate between application, control and data planes, it is possible to address and control each node in a network to optimise global performance. This also removes the application specific nature of many WSNs [5], [6] allowing a single WSN to fit multiple purposes.

However, SDN was originally intended for wired networks with control messages orthogonal to the data, with a static topology and over a reliable channel [3]. When applied to wireless networks, orthogonality is achieved by redirecting the control plane over a wired backbone or over a secondary wireless interface using a technology that is able to handle the increased channel occupancy and energy demand while still guaranteeing reliable message delivery [4]. As WSN networks have limited range, capacity, reliability and energy, with mobile nodes using a single network interface, this approach is usually not possible and as such limits the practical advantages of the SDN paradigm as it constraints the control traffic to the same impediments of the data messages [7]. Figure 1 shows the architecture of an SDN-controlled WSN in which both control and data travel over wireless channels.

Bluetooth Low Energy (BLE) has become the de-facto standard for indoor low-power and low-datarate WSN [8]. It has seen major adoption in both the smart home [9] and e-Health [10] markets as it provides good indoor coverage and compatibility with a vast array of off-the-shelf devices. Recent efforts have been carried out to introduce mesh networking capabilities to BLE, thus allowing to expand the range of the BLE network without incurring in the costs of additional hardware of the traditional Bluetooth cellular topology [11]. A recent survey has shown that the performance of BLE mesh solutions varies widely depending on the protocol implementation and design choices, especially on the routing strategy [11] while the work in [12] has shown that there is strong dependency between the design choices and the actual implementation of routing BLE mesh networks. However, as of today BLE mesh solutions suffer from the aforementioned classical WSN problems and, to the best of the authors' knowledge, there is no proposal yet in the State of the Art (SoA) to implement the SDN paradigm in BLE mesh.

The main contribution of this work is then the creation of a proof-of-concept SDN technology applied to a real-life BLE mesh testbed. The behaviour of the developed system is studied and it is shown how a network can recover from a congestion which would have been fatal without the control infrastructure in place.

The remainder of this paper is structured as follows. Section II motivates the current need of an SDN approach in BLE mesh and introduces the SoA of SDN for WSN. Section III presents the hybrid BLE mesh design and discusses how this solution enables SDN implementation. Section IV introduces the construction of the BLE testbed. Section V discusses the measurements and results of the proof-of-concept and, finally, Section VI concludes this work.

## II. MOTIVATION AND RELATED WORK

### A. Motivation for SDN for BLE Mesh

BLE makes use of two different packet structures: broadcast packets, transmitted over **three dedicated advertisement channels** and which do not require the establishment of a connection between nodes nor acknowledgement, and connection packets which are transmitted over the remaining **37 data channels** and require the creation and maintenance of a point-to-point link between two nodes. The BLE mesh protocols then make use of either *flooding* strategies, where the data packets are broadcasted on the advertisement channels and there is no guaranteed QoS (as in BLEMesh, the protocol actually present in the Bluetooth 5 standard [13]) or they make use of *routed messages* over the data channels, where a path can be constructed between a source and a destination and has to be maintained [14], [15]. The trade-offs between the these two different types of meshing protocols have been explained in [16], [17].

The separation between advertising and data channels in the BLE standard allows then a unique opportunity to bring an SDN approach to IoT WSN networks by using the data channels to carry data between the nodes and by using the advertisement channels to spread control information from the nodes to a centralised controller and to direct node behaviour. This way the network is efficiently split into two slices, one for control and one for data which do not share spectrum and path properties as the messages use different frequencies and travel through different paths. Each slice can have, by design, different requirements and specifications: low latency slices for urgent data, control messages or to serve mobile nodes, and reliable slices for long lived data streams that require reliability and energy efficiency. Note that the current implementation of BLEMesh, present in the Bluetooth standard, relies on mere flooding to propagate the packets over the network. As such, nodes need to be always scanning (equivalent to a duty cycle of 100%), which increases their energy consumption, and additionally there is no randomization in their transmission timers, which may lead to increased collisions. In fact, the role of the Friend Node in BLE [18] is necessary to avoid exhaustion of batteries. This is a trivial solution that does not allow fine tuning of any parameter depending on application needs. With the proposed SDN framework, energy consumption, network size, packet delivery ratio and end-to-end delay can be fine tuned to allow different types of data streams in the network. Moreover, it is able to react dynamically to varying network conditions and automatically solve network congestion to ensure that application needs are met, as shown in Section V.

### B. SDN for WSN

Enabling SDN operation in wireless networks has attracted much interest in the scientific community, but concrete implementations of SDN for WSN are still novel. Recent surveys on WSN-SDN solutions and their implementation challenges in [4], [19], [20]. The vast majority of SDN for WSN proposals make use of OpenFlow as a communication protocol and are based on IEEE 802.15.4 functionalities [5], [6]. Some other works propose IP-based WSN using 6LoWPAN [21], but require additional compression to not overload the resource constrained nodes. However, to the best of the authors' knowledge, there is no current proposal for SDN based on BLE.

## III. PROPOSED SDN IMPLEMENTATION FOR BLE MESH

### A. Brief SDN overview

In SDN, the network control is decoupled from the actual hardware nodes generating and forwarding the data. This allows to implement control logic which is independent of the physical and topological condition of the network and is closely linked to the application and end-to-end requirements [22]. Generally, SDN can be described by a three-layer structure: at the bottom is the *Data Plane* (DP), containing all the hardware nodes such as sensors, routers, and actuators; on the top is the *Application Plane* (AP), in which the application specifications and on-demand requirements can be monitored; and, finally, bridging the two in the middle is the *Control Plane* (CP), which adapts network behaviour and controls the DP with respect to the requirements coming from the application [19].

Each layer intercommunicates making use of APIs. Specifically, the **Southbound API** carries the control signals from the CP to the data forwarding nodes in the DP (e.g. routing tables, bandwidth allocation, channel selection strategy, etc.) and reports back the behaviour of each node from the DP to the CP. The **Northbound API** carries the end-to-end requirements from the AP to the CP and, vice versa, it carries the abstracted information of the network to the application [20]. Finally, the **East- and West-bound APIs** connect different modules of the CP, in case this one would be distributed over multiple physical entities.

*B. Overview of the proposed SDN design for BLE Mesh*

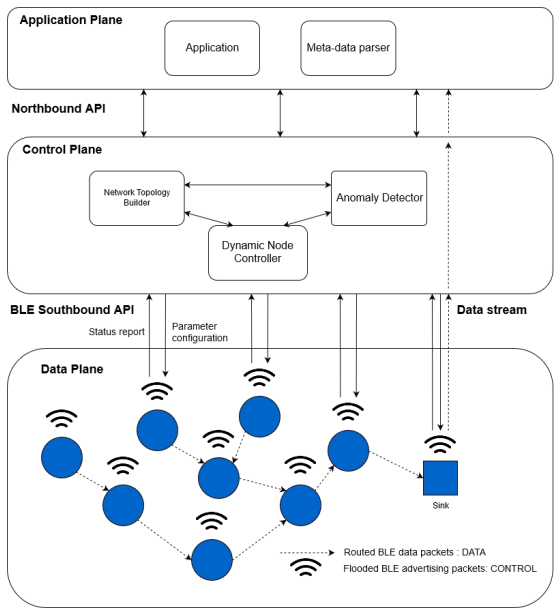

Fig. 2: Proposed SDN implementation for BLE mesh networks where the BLE data packets are routed through intermediate connected nodes and the control packets of the Southbound API are broadcasted using advertising packets.

The structure of the proposed framework is depicted in Figure 2 and the three SDN layers with their corresponding APIs will now be exposed. Note that both DP and Southbound API traffic flows operate in the wireless domain but the former makes use of connected data packets while the latter makes use of broadcasted advertising packets.

| Control | |
|---|---|
| Field | Comment |
| *Node_ID* | Node identifier |
| Device Type | [Sink/Static/Dynamic] |
| Master | (Node_ID, RSSI, CI, Queue) |
| Slave_1 | (Node_ID, RSSI, CI, Queue) |
| Slave_2 | (Node_ID, RSSI, CI, Queue) |
| Slave_3 | (Node_ID, RSSI, CI, Queue) |
| Battery | Battery level |
| Tx_Power | |
| Timestamp | |

TABLE I: The *uplink* Southbound API status report messages contains BLE-specific fields that enable the CP to infer the status of the DP.

Orthogonality is ensured since the two BLE operations use different paths, channels, timeslots and state machines.

*1) Resource constrained Data Plane:* The DP contains the wireless BLE nodes, responsible for sensing the environment, forwarding data packets to a sink and reporting networking metrics to the CP.

Data packets generated by the nodes represent a continuous stream of sensitive information where reliable packet delivery needs to be guaranteed. The DP then benefits from a routing scheme where BLE master-slave connections are established between neighbor nodes: message reliability against channel impairments is ensured by retransmissions and frequency hopping over BLE data channels, while the synchronized sleeping and waking up patterns between connected devices ensure a lower energy consumption and the possibility of optimizing this Connection Interval for the traffic of the network, so end-to-end delay is not excessively increased. Data packets will then travel over multiple regular BLE connections until they arrive at the sink.

Note that the master-slave mechanism results in a data mesh network with tree topology, with the sink on top of it.

*2) BLE Southbound API:* The Southbound API delivers periodic status reporting from the DP to the CP on each node's behaviour and characteristics and also also allows to send parameter configuration messages from the CP to the DP. In order to ensure orthogonality with the data stream and to disconnect the forwarding of control packets with the topology of the network, the Southbound API makes use of broadcasted BLE advertising packets. The rate of broadcasting can be optimized for each node so the reporting is done often enough to keep track of the activity of the node but without incurring in a high additional energy consumption and message overhead. Additionally, a flooding mechanism is used where the nodes rebroadcast control packets of their neighbors as a way of ensuring path diversity and transmission reliability.

Table I shows the metrics reported from DP to CP assuming each BLE node has one master and three slaves as an example. The control information coming from the DP summarises the condition of a node in the time elapsed since the last message. The BLE node identifies itself by including the Node ID with each message and it specifies whether it is a sink, a static or a mobile node. The node then appends the basic communication statistics between itself and its master, particularly, it sends the master's ID, the received RSSI, the Connection Interval and the average queue occupancy. The very same information is repeated for all the connected slaves as well. The node then appends its battery level, transmit power and timestamps the message. This information contains both the state of the node and average traffic information for all of the node's connections.

When control information comes from the CP downwards, on the other hand, the message contains configuration parameters for a specific node, as exemplified in Table II. This over the air parameter configuration is used to individually taylor the performance of the nodes in order to adapt to changing network conditions.

| Control | |
|---|---|
| Field | Comment |
| *Node_ID* | Node identifier |
| Command | Change X parameter towards C connection |
| X parameter | Connection Interval |
| C connection | Slave_3 |
| Timestamp | |

TABLE II: A possible *downlink* Southbound API parameter configuration message from CP to DP.

*3) Control Plane:* The CP contains all the control logic of the BLE mesh network. Such logic could run either on the sink nodes or on a common point of arrival for all the data in case of distributed networks [7]. The implemented control logic resides, for this work, on a server where all the BLE nodes transmit data to. The controller is responsible for changing the connection parameters of each node to suit dynamically changing environment and application conditions. The controllable parameters in the testbed are the BLE connection interval, the transmission power, the buffer size and the number of slave connections.

Manipulating any of these variables alters the characteristics of the link between two nodes, for example, by increasing or decreasing the available bandwidth (by, in turn, choosing a lower or higher connection interval) paying thus an energy cost. This single link control has then end-to-end repercussions which impact the overall reliability and can be witnessed only at the AP.

Another extremely important task of the CP is then to provide the AP with an abstracted but functional model of the network. This entails information on the condition of the nodes and of the overall network topology; while the DP does, in fact, provide the actual application data, it is possible to extract networking information directly form the data packets' meta information. This information needs to be conveyed to the CP in which then three main actors reside: the Network Topology Builder, the Anomaly Detector and the Dynamic Node Controller.

The **Network Topology Builder** is responsible for determining the shape of the network from control packets received by each node via the Southbound API. The **Anomaly Detector** makes use of meta-data collected in the AP and fed back via the Northbound API to determine whether the expected amount of data is received from each node and, if not, whether this is due to sensing errors or to congestion. Finally, the **Dynamic Node Controller** is responsible for manipulating each node in order to keep reliability high and avoid congestions. All of these functionalities will be expanded in Section III-C.

| Northbound message | |
|---|---|
| Field | Comment |
| Node ID | ID of the node transmitting the message received at the sink |
| PID | Packet identifier of that specific received packet |
| DATA | Number of payload bits contained in the packet |
| HOP | Hop count between source node and destination |
| Timestamp_RX | Time at which the packet has been received at the sink |

TABLE III: Meta-data extracted at the the AP and sent down to the CP via the Northbound API.

*4) Northbound API:* The Northbound API acts as the communication layer between the CP and AP. Specifically for low-power WSNs it is important to keep communication overhead low; this entails that it is imperative to learn as much as possible of the behaviour of the network from the application data packets; for this reason the proposed solutions make use of passive diagnosis methods in which no probing is necessary but only application packets received at the sink are considered. This is obtained by parsing the meta-data associated to each data packet and feeding it back from the AP to the CP. This closes the loop and allows for real-time monitoring of end-to-end performance, which is totally absent from current low energy WSNs.

An example of the Northbound API can be seen in Table III. It is important to point out that the Northbound API does not make use of BLE but it is, in the current work, a logical channel as the AP and CP reside both on a server.

*5) Application plane:* Finally, a specific **Application Data Parser** is used at the AP to extract useful meta-data from application packets already received. This allows to gather important end-to-end performance metrics without burdening the network with intense control communication overhead. For each application packet received then, the parser extracts the information present in Table III and sends it down to the CP.

## C. Detailed functionalities of the Control Plane

The key elements of SDN operation are the agents composing the CP. Brief but clear explanations of their structures are given below to understand the complete control chain.

The **Network Topology Builder** builds a graph of the connections in the BLE network using the Southbound messages of Table I. As in the current implementation each BLE node can have only one master, it is then trivial to build a directed graph with the control information obtained. Furthermore, the topology builder can visualize the graph and perform operations to determine the distance between two nodes and the shortest path. Figure 3 shows an example of graph showing the nodes represented by circles containing their Node IDs, with the sink node being on top of the tree.

The **Anomaly Detector**, on the other hand, is built based on the probabilistic passive diagnosis model proposed by

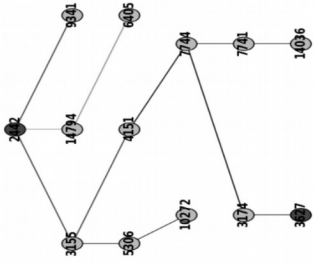

Fig. 3: An example of network graph showing the BLE nodes represented by circles containing their Node IDs.

Liu et al. in [23]. The passive diagnosis work in [23] uses a Bayesian Network to infer whether a node has failed and differentiates between two possible failures: *sensing*, i.e. a node was supposed to sense data but it is not doing it, or *communication*, i.e. a node was supposed to send data to the next hop but somehow it has failed to do so. The Anomaly Detector requires the path between each node and its sink and the amount of data generated by each node. The former is obtained directly by the Network Topology Builder which keeps the anomaly detector up do date with the current topology; the latter is obtained directly from the meta-data sent through the Northbound API. The Anomaly Detector is aware of the amount of traffic generated by each node received at the sink (Node ID, PID, DATA and Timestamp RX presented in Table III contain all the necessary information). The Anomaly Detector then produces continuous reports for each node containing information on faulty node sensing or communication probabilities (100 % probability means that either the node or the link are down).

Finally, the **Dynamic Node Controller** receives the reports from the Anomaly Detector and sends, through he Southbound API, configuration settings for the BLE nodes. If an anomaly report contains a probability of a sensing anomaly above predetermined threshold the node is considered faulty (e.g. the node has died because the battery was faulty) and a network operator might be notified. If, on the other hand, the anomaly report contains high probability that a communication anomaly for a specific node, this entails that the nodes are generating the correct amount of data but the data link between the node present in the report and the next hop is not able to carry the packets properly. This can be due to a lack of connection entirely (e.g. the radio is broken) or to a congestion (e.g. a node has to relay too much information and is unable to send all the data with the current configuration). In case there is a congestion, the node configuration parameters can be manipulated for any particular node via the Southbound API to alleviate or remove the condition. It is important to point out that there isn't, as of today, a direct method to link the anomaly probability values with an actual network condition in a general way. Different network topologies and traffic types will give rise to different dynamics and, as such, the probabilities will be different for different faults in different networks and the system would need to be calibrated effectively in order to select appropriate thresholds for appropriate actions (e.g. differentiate between a congestion and just a small processing delay).

## IV. TESTBED IMPLEMENTATION

This section briefly introduces the implementation of the proposed SDN framework in a hybrid BLE mesh testbed.

### A. Architecture and hardware

The testbed consists of a set of BLE nodes, single-board computers, a network switch and a PC server, as shown in Figure 4. The BLE nodes create a mesh network where the DP is located. Every single-board computer has four nodes connected, so the received packets are parsed and sent to a PC server. The CP is then split between the single-board computers and the server, where the AP resides.

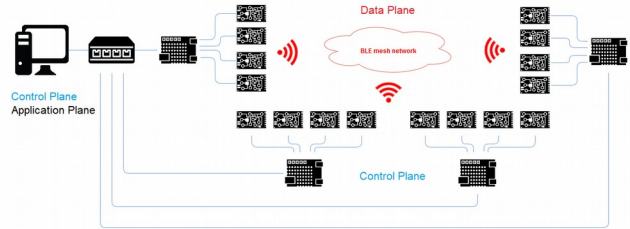

Fig. 4: Architecture of the testbed with Data Plane (BLE mesh network), Control Plane (single-board computers and PC server) and Application Plane (server).

The BLE nodes selected are the Nordic Semiconductor nRF52 development boards [24]. They are flashed with a BLE stack and the routing and flooding mesh protocols for the DP and the Soutbound API are developed on top of it. Every four nRF52s are connected to a single-board computer through their USB connection to power them and to additionally receive commands and print their output.

Hardkernel Odroid-C2 are deployed as single-board computers [25], which establish a UART session with each of connected nodes. When a node receives a packet it sends it over the UART and the Odroid parses it and logs it on the server through the Ethernet connection.

The PC server is the central point and the user interface of the testbed. It receives and stores all data generated by the nodes, while providing visual information about the state of the mesh network.

### B. Software

All DP activity is carried out by the BLE nodes, where the BLE stack and mesh protocols are executed. The Odroids act as physical controllers for the BLE nodes: they offer terminal connection and debugging functionalities. Additionally, they become gateways for the BLE sink nodes, parsing the meta-data from the DP packets and sending it to the server where the three control agents (Anomaly Detector, Topology Builder and Dynamic Node Controller), developed in python, are running. Finally, the server also provides several applications for monitoring in real time the state of the network or logging all packets received into a database.

## V. PROOF OF CONCEPT

This section presents a proof of concept for the proposed SDN implementation where the resource constrained DP suffers from a congestion situation and, thanks to the control messages sent over the Southbound API, the three CP agents are able to automatically detect and solve this issue.

A 12 node mesh network is set up, with Table IV showing the values of the main parameters set in the nodes unless otherwise stated.

| Connection Interval | 300 ms |
|---|---|
| Transmission Power | 0 dBm |
| Buffer size | 25 packets |
| Packet size | 20 bytes |
| Data packet generation period | 10 s |
| Control packet generation period | 30 s |

TABLE IV: Default node parameters.

Figure 5 shows the topology of the network, where every node is only allowed to have a master and a slave connection. Albeit simple, this topology represents the worst case scenario in terms of congestion: as each node generates own data packets and routes them towards the sink node $N_{12}$, nodes closer to the sink need to deliver a higher load of traffic in order to satisfy the network demands.

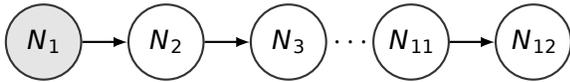

Fig. 5: Single line multihop data mesh network. Nodes transmit their own data at a constant rate and additionally forward packets to the sink, except $N_1$ which has a variable packet generation rate.

The data rate is constant for all nodes in the network with the exception of $N_1$, which eventually starts generating 2 data packets per second (a rate 20 times higher than in the rest of nodes). The network then transitions from a **Steady State** into a **High Demand State**, where the nodes do not have enough bandwidth to support this data rate. Figure 6 depicts the instantaneous buffer occupancy in the network (averaged for all nodes) for every correctly received packet at the sink and originated by $N_1$. As the connection interval of 300 ms is not capable of delivering packets at this rate, the buffers from all intermediate nodes start to populate. This situation continues until all buffers are full and the network transitions into a **Congestion** state. Without the proposed SDN framework this situation would not be detected nor solved, staying permanently in this state. However, the Anomaly Detector controller is able to detect that packets are being lost and extract the nodes and links responsible for this congestion. The Dynamic Node Controller then reconfigures the connection interval of all links so the high demand can be supported. This reconfiguration is done over the air and these packets are sent over the Southbound API using advertising packets over the flooding mesh protocol, thus avoiding the congested links. Once the reconfiguration is automatically done, the network is able to recover and return to the **Regular Performance** state.

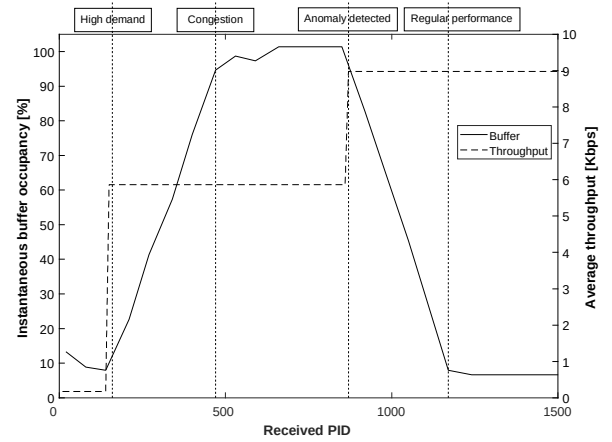

Fig. 6: Instantaneous buffer occupancy for all nodes and average throughput of the network.

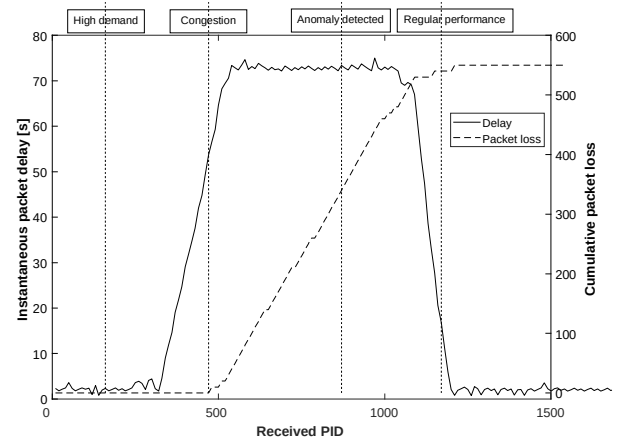

Fig. 7: Instantaneous packet delay and cumulative packet losses over time during the network experiment.

Additionally, Figure 6 shows the average throughput of the network. In the **Steady State** all nodes transmit a data packet every 10 seconds, which translates into an average throughput of $R = 176$ bps. Then, in the **Congested State** the nodes transmit at their maximum possible rate for the connection interval used, which results in a throughput of $R = 5.8$ Kbps. Note, however, that this throughput is not enough to satisfy the demands of the network. This is only possible once the anomaly is detected and solved, with the connection intervals adapted so all traffic can be delivered to the sink, resulting in a throughput of $R = 9$ Kbps once the final **Regular Performance** state is reached.

Figure 7 depicts the instantaneous packet delay for every received packet and the cumulative packet loss. Following the same events described above, the packet delay ramps up from the **Steady State** once the buffers are populated and saturates to its maximum value during the **Congestion** state. At this point, every packet that is not dropped due to buffer overflow needs to travel through all buffers from all nodes in the network, which translates into a steady delay of 75 s. Note that this curve is not aligned with the four states marked precisely due to the delay of the received packets. Again, when the anomaly is solved and the **Regular performance** state is reached the buffers are emptied and the delay decreases back to its initial low value. Regarding the cumulative packet loss, no transmissions are corrupted due to the channel and only packets dropped due to buffer overflow will increase its value. This occurs in the **Congestion state** and no more packets are lost once the anomaly is detected and the Dynamic Node Controller re-configures the connection interval of the nodes.

## VI. CONCLUSIONS

This paper proposes a novel proof-of-concept to bring SDN solutions to BLE mesh networks composed by mobile and resource constrained nodes. Control and Data planes are split orthogonally but are both using the wireless medium. This allows to efficiently split the network into two slices: one for low latency, unreliable and high energy consuming information streams and another one for applications that require reliability and energy efficiency.

The SDN for BLE framework is introduced. A description of the architecture is given, showcasing the main elements of each plane and how these are interconnected using APIs. The three functionalities developed for the Control Plane are described: the Topology Builder, the Anomaly Detector and the Dynamic Node Controller.

Finally, the proposed SDN framework is implemented in a BLE testbed and a proof-of-concept scenario is presented, where changing network conditions lead to a congestion situation. The framework is able to automatically detect this issue and solve it by identifying the nodes responsible for such problem and tuning their parameters.